\documentclass[12pt]{article}
\usepackage{graphicx}
\usepackage{subfigure}
\usepackage[latin1]{inputenc}
\pdfoutput=1
\usepackage[english]{babel}
\usepackage{amsmath}	
\usepackage{amssymb}	

\def\be{\begin{equation}}
\def\ee{\end{equation}}

\def\bea{\begin{eqnarray}}
\def\eea{\end{eqnarray}}

\newcommand{\OOO}{\mathcal O}
\newcommand{\mat}[4]{\left(\begin{array}{cc} #1 & #2 \\
    #3 & #4 \end{array}\right)}

\newcommand{\vect}[2]{\left(\begin{array}{c} #1 \\
    #2\end{array}\right)}

\def\beq{\begin{equation}}
\def\eeq{\end{equation}}

\begin{document}

\vspace*{-1cm}
\thispagestyle{empty}
\begin{flushright}
LMU-ASC 13/13\\
LPTENS 13/06
\end{flushright}
\vspace*{0.2cm}

\begin{center}
{\Large 
{\bf Fusion of  Critical Defect Lines in the 2D Ising Model}}
\vspace{2.0cm}

{\large C.~Bachas${}^{\, \sharp}$}%
, \hspace*{0.1mm}  \hspace*{0.2cm} {\large I.~Brunner${}^{\, \flat, \, c}$}%
  \hspace*{0.3cm} {\large and}  \hspace*{0.2cm} {\large D.~Roggenkamp${}^{\, \natural}$}%
\vspace*{0.5cm}

${}^\sharp$ Laboratoire de Physique Th\'eorique de l'Ecole 
Normale Sup\'erieure 
\footnote{Unit\'e mixte de recherche (UMR 8549)
du CNRS  et de l'ENS, associ\'ee \`a l'Universit\'e  Pierre et Marie Curie et aux 
f\'ed\'erations de recherche 
FR684  et FR2687.}
\\
 24 rue Lhomond, 75231 Paris cedex, France\\
\vspace*{0.4cm}
   
 ${}^\flat$ 
Arnold Sommerfeld Center, Ludwig Maximilians Universit\"at \\
Theresienstra\ss{}e  37, 80333 M\"unchen, Germany \\
\vspace*{0.4cm}

${}^c$  
Excellence Cluster Universe, Technische Universit\"at   
M\"unchen \\
Boltzmannstra\ss{}e  2, 85748 Garching, Germany \\
 
\vspace*{0.4cm} 

${}^{\, \natural}$ 
Institute for Theoretical Physics,                             
University of Heidelberg\\
Philosophenweg 19, 
69120 Heidelberg, 
Germany
\vspace*{2cm}

{\bf Abstract}
\end{center}

  Two  defect lines separated by a distance $\delta$ look  
 from much larger  distances  like a single defect. In the critical theory, when all scales 
  are large compared to the cutoff scale, this  fusion of defect lines is universal. 
   We calculate the universal fusion rule  in the critical 2D Ising model  and show  that it is given by the Verlinde algebra
   of primary fields, combined with group   multiplication in $O(1,1)/Z_2$.
Fusion is in general singular  and  requires the subtraction of a divergent Casimir  energy.

\newpage

\section{Introduction and Summary}

     Ever since Onsager's celebrated solution \cite{Onsager:1943jn}, the two-dimensional Ising model has been the prototype
for the study of  second-order phase transitions.   The model also exhibits critical behavior  
 on  boundaries  \cite{Binder},  and  on  defect lines. The latter   have been  analyzed using  both integrability (\cite{Henkel:1988vh,Abraham:1989cz} 
 and references therein) 
 and conformal field theory \cite{Oshikawa:1996ww,Oshikawa:1996dj} techniques. It  has been found, in particular, that the 
  critical behavior of defect lines  is captured by the three continuous
 families given in  table \ref{table1}.
   
 \smallskip
 
 The purpose of the present  note is to  compute  the fusion algebra \cite{Bachas:2007td}  of these conformal  defects: when two of them are placed parallel to each other,  they  fuse to another such defect line in the limit of zero separation. The process
  is in general singular, and requires the subtraction
 of a divergent self-energy.

 It turns out that the resulting fusion algebra takes a simple form in the fermionic representation of the Ising model. There, defect lines are parametrized by a gluing matrix $\Lambda\equiv -\Lambda\in O(1,1)/\mathbb{Z}_2$ of the fermions, which has to be an element of the Lorentz group in 1+1 dimensions (modulo its center), and by  an Ising primary  $a\in \{\boldsymbol{1}$, $\boldsymbol{\epsilon}, \boldsymbol{\sigma} \}$. Defect fusion then reduces to a combination of multiplication in the Lorentz group,  and multiplication in the Verlinde algebra of the Ising model ($ \boldsymbol{1} \times a = a$, $\boldsymbol{\epsilon}\times \boldsymbol{\epsilon}= \boldsymbol{1}$,
 $\boldsymbol{\epsilon}\times \boldsymbol{\sigma}= \boldsymbol{\sigma}$ and 
 $\boldsymbol{\sigma}\times \boldsymbol{\sigma}= \boldsymbol{1} + \boldsymbol{\epsilon}$, see {\it e.g.} \cite{Verlinde:1988sn}).
 Explicitly, defects associated to $(a,\Lambda)$ and $(a^\prime,\Lambda^\prime)$ fuse according to 
  \be\label{Fusion}
  (a, \Lambda) \star (a^\prime, \Lambda^\prime) = (a\times a^\prime, \Lambda \Lambda^\prime)\,.  
  \ee
  For the special subclass of defects with diagonal gluing matrix $\Lambda$ fusion was previously obtained in  \cite{Petkova:2000ip}. These are {\it topological} defects and their fusion is non-singular. 
Here,  using the results of  \cite{Bachas:2012bj}, 
we  will derive fusion of general conformal defect lines in the Ising model, {\it i.e.} of all defects obtained by
marginal deformations of the topological  defect lines.

\medskip
  
\begin{table}[h]\label{table1}
\begin{center}  
\begin{tabular}{|r|r|r|}
\hline
Spin-chain defect    \ \  \ \ \ \ \ \  &  $\mathbb{Z}_2$-orbifold  boundary\  \ &  Fermionic \ \ \   \\
\hline\hline
\, &  \,   & \,   \\
ferromagnetic, $b\in (0, \infty)$\  \ \   &  Dirichlet, $\phi_0\in(0, \pi/2)$\,  & $\mathbf{1}$, ${\rm det}\Lambda = 1$   \ \ \\
\, &  \,   & \,   \\
 anti-ferromagnetic,  $b\in (-\infty, 0)$  &  Dirichlet, $\phi_0\in( \pi/2, \pi)$\,   & $\boldsymbol{\epsilon}$, ${\rm det}\Lambda = 1$   \ \ \\
   \, &  \,   & \,   \\
 order-disorder,  $\tilde b\in (0, \infty)$\  \ \  
  &   Neumann, $\tilde\phi_0\in(0, \pi/2)$  &  $\boldsymbol{\sigma}$, ${\rm det}\Lambda = -1$   \\
 \, &  \,   & \,   \\
\hline
\end{tabular}
\caption{\footnotesize Universality classes of  defect lines in the Ising model. The left column 
gives the natural  parametrization in terms of Ising spins. The central one the 
corresponding boundary states in the  $c=1$ CFT. Finally the right column gives the  parametrization  in terms of gluing 
matrices for the fermion fields and Ising primaries. } 
\end{center}
\end{table}

   The  Ising model on a square lattice with an integrable,  ferromagnetic or anti-ferromagnetic defect 
   line has  the energy-to-temperature ratio
\be\label{Isingc}
{{\cal E}\over T} = - \sum_{i, j}  ( K_1 \sigma_{i,j}\sigma_{i+1,j}  + K_2 \sigma_{i,j}\sigma_{i,j+1} ) +   (1-b) K_1 \sum_j  \sigma_{0,j}\sigma_{1,j}
\ee
where  $\sigma_{i,j} =\pm 1$ are the spin variables,  and 
${\rm sinh}(2K_1)\, {\rm sinh}(2K_2) = 1$ in order for the bulk theory to be critical. 
Couplings along the (vertical) defect line   are rescaled by a factor $b$, which parametrizes marginal deformations
of the defect. These defects correspond to conformal defect lines specified by Ising primaries $a\in\{\boldsymbol{1},\boldsymbol{\epsilon}\}$ and fermion-gluing matrices 
\be\label{det1gluing}
\Lambda=  \mat{{\rm cosh}\gamma}{{\rm sinh}\gamma}{{\rm sinh}\gamma}{{\rm cosh}\gamma}\ 
\ee 
of determinant 1, {\it c.f.} table \ref{table1}.
As we will see,  the 
 relation between  the defect strength  $b$ and the hyperbolic angle $\gamma$ of the gluing matrix is given by
 \bea\label{relnt}
  \gamma =  {\rm log}  \Bigl\vert { {\rm tanh}(b K_1) \over {\rm tanh}(K_1)} \Bigr \vert \ . 
\eea
Since the Lorentz matrices \eqref{det1gluing} multiply by adding the hyperbolic angles $\gamma$,
the fusion of two defects with couplings $b$ and $b^\prime$ results in a defect 
with coupling  $b^{\prime\prime}$, where 
 \be\label{rule}
{\rm tanh}( b^{\prime\prime} K_1)\,  {\rm tanh}(K_1)  =  {\rm tanh}(b K_1)\, {\rm tanh}(  b^\prime K_1)\ . 
 \ee
 Notice that we wrote the  fusion rule without the absolute values coming from \eqref{relnt}. 
Indeed, the signs of the defect strengths combine multiplicatively, in accordance with  the $Z_2$ algebra
 of the Ising primaries $\boldsymbol{1}$ and $\boldsymbol{\epsilon}$.

 \smallskip
 The Ising model also features order-disorder defects which are  obtained by performing a 
 duality transformation on one side of the (anti-)ferromagnetic defect lines.  As detailed in table \ref{table1}, these  
  correspond to  the conformal  defects with $a=\boldsymbol{\sigma}$ and fermion
gluing matrix
\be\label{det-1gluing}
\tilde{\Lambda}=  \mat{{\rm cosh}\tilde\gamma}{-{\rm sinh}\tilde\gamma}{{\rm sinh}\tilde\gamma}{-{\rm cosh}\tilde\gamma}\ 
\ee 
of determinant $-1$. The microscopic realization of these  defect lines is most simple  in the strongly-anisotropic limit of the critical
Ising model, $K_1\to 0$  (which implies $K_2\to\infty$). In this limit one has
\be
e^\gamma = \vert b\vert   \qquad {\rm and} \qquad e^{\tilde\gamma} =   \tilde b  \ , 
\ee
where  $\tilde b$ is the coupling strength  of the order-disorder defect.\footnote{Performing the 
duality transformation on a ferromagnetic defect with coupling $b$ and an anti-ferromagnetic one with coupling 
 $-b$ yields the same order-disorder defect. Thus, one may restrict  the range of the parameter $\tilde b$ of the order-disorder defects to $(0,\infty)$.}
\smallskip

The fusion of two order-disorder defects turns out to produce the sum of a ferromagnetic and an anti-ferromagnetic defect 
of the same absolute strength  $\vert b^{\prime\prime}\vert$.  Since the Lorentz matrices \eqref{det-1gluing} multiply by subtracting the
 hyperbolic angles, one finds
 \be
   \vert b^{\prime\prime} \vert    =  \tilde b  /  \tilde b^\prime \ . 
 \ee
Notice that  two order-disorder defects only commute if they are identical. 
Likewise the fusion of an (anti-)ferromagnetic with an order-disorder defect line produces an order-disorder defect line with coupling
 \be
   \tilde b^{\prime\prime}      =  \vert b  \vert  \,  \tilde b^\prime \qquad {\rm or}\qquad 
    \tilde b^{\prime\prime}      =    \tilde b  / \vert b^\prime  \vert  \ ,  
 \ee
depending on whether the (anti-)ferromagnetic defect is fused from the left  or the right.
 Defect fusion is non-commutative.
\smallskip 
  
 The above rules for fusion of defect lines   are the main results of this letter.  
 They are summarized by the master formula \eqref{Fusion}. 
  We should stress that although  the fusion algebra is universal,   the   
 parametrization of the critical  lines of defects is not.  
  In particular, relation \eqref{relnt} depends on the non-universal constant $K_1$.  
  Note also that the stability of the order-disorder defects  is ensured by   a $\mathbb{Z}_2\times \mathbb{Z}_2$ symmetry, 
  which reflects separately the spins on the two sides of the defect line, 
    whereas
  the more stable  (anti-)ferromagnetic defects only preserve  the  diagonal $\mathbb{Z}_2$  \cite{Oshikawa:1996dj}.


 \section{Fusion of Conformal Defects}
 
  Figure 1 illustrates the physical meaning of fusion of conformal defects: 
 we consider two defect lines  ${\cal D}$ and ${\cal D}^\prime$ 
    separated by a distance $\delta$, and let $x$ be the  typical  (horizontal) scale  at which  the system is probed.
    For   $x \gg \delta$  the system  flows  to an effective defect line ${\cal D}_{\rm eff}$, but   in general this ${\cal D}_{\rm eff}$ will depend
    on   $\delta$ and on  the precise microscopic realization  of the defects ${\cal D}$ and ${\cal D}^\prime$.  
    Put differently, the composition  of   two defects for finite $\delta$  is not universal. If, however, 
     $\delta$ is also large compared to the lattice spacing $\Delta$,    then one expects the fusion to only depend  on the universality classes of 
     ${\cal D}$ and ${\cal D}^\prime$. This  universal composition  rule can be calculated in conformal theory.

\begin{figure}[ht]
\centering
\includegraphics[width=0.92\textwidth]{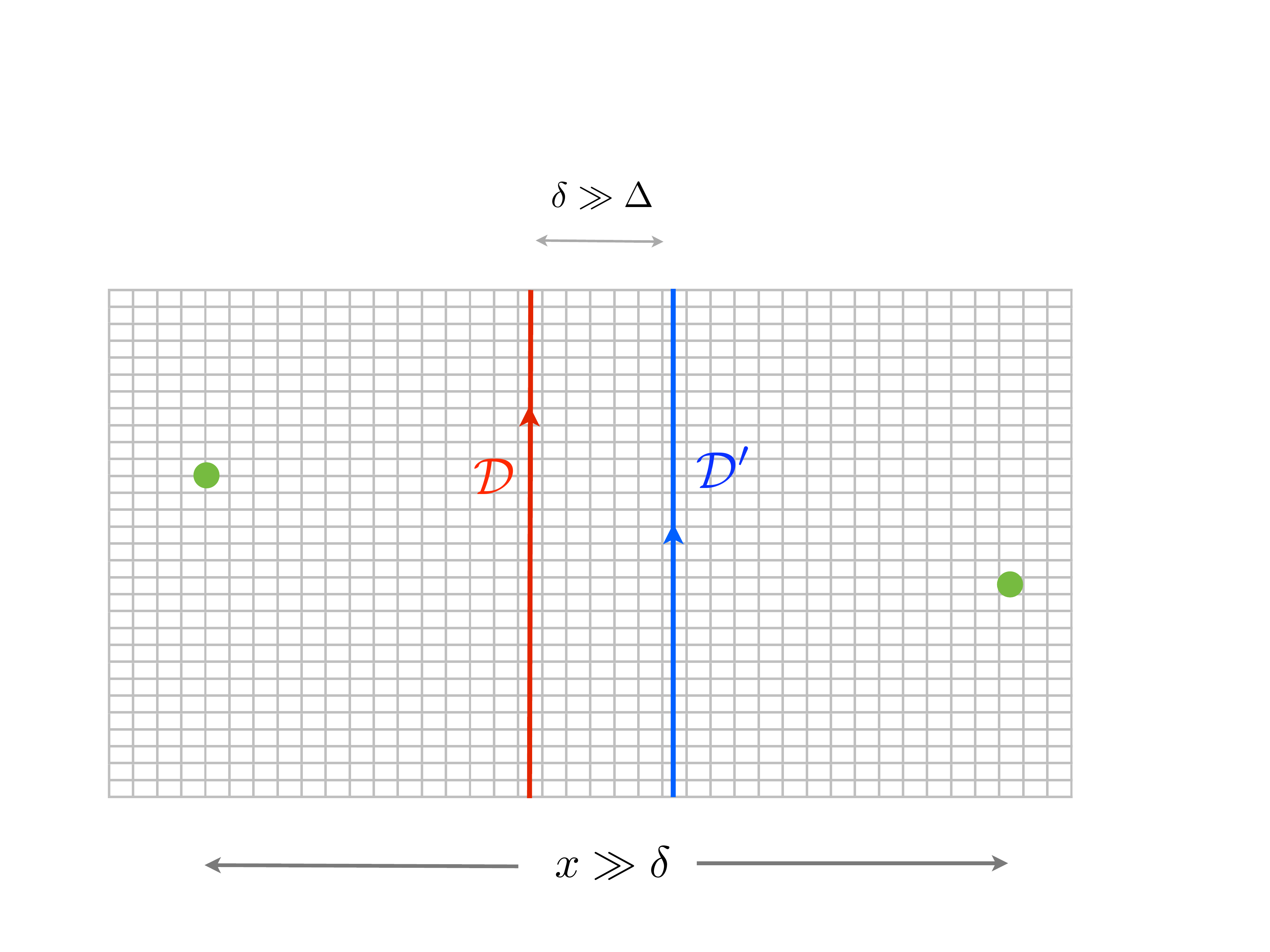} 
\caption{\footnotesize The two-defect system  discussed in the text. The green dots
  are arguments of a typical two-point function  at a horizontal scale   $x\gg \delta$. 
  The fusion product   ${\cal D}\star {\cal D}^{\prime}$ gives an effective description of this system
  in the limit where  $\delta$  is very  large compared to the lattice spacing $\Delta$. Only in this limit is  fusion  universal.
   }\label{fig:fuse}
\end{figure}

\smallskip    
      
       To perform the calculation, one  may quantize  the CFT by compactifying  the defect line on a circle and treating the normal direction
        as time. The defect is then described by a formal operator  which acts on the space
         of states of the CFT on the circle. This generalizes the technical device of  boundary state \cite{Callan:1987px}
          to defect lines. 
         The action of two coincident defects is given by the product of the corresponding operators,
         but this is in general singular and requires regularization and renormalization.  
         
         \smallskip
         A simple example, that of $U(1)^2$ invariant defects
          in the $c=1$ CFT \cite{Bachas:2001vj},  has been  worked out in detail in reference  \cite{Bachas:2007td}.  In this case a single subtraction
           of a divergent Casimir energy is sufficient to render the result finite,\footnote{Even this is not needed in the case of 
           unbroken supersymmetry, as in
           the examples considered in   \cite{Bachas:2012bj,Mikhailov:2007eg,Benichou:2010ts}.    
           }
           so one defines 
           \be\label{regularize}
            {\cal D}\star  {\cal D}^\prime =  {\rm lim}_{\delta\to 0}\  [  e^{- C/\delta}\, {\cal D} \,  e^{-\delta\,{\cal H}} \,  {\cal D}^\prime ]\ , 
           \ee
         where ${\cal H}$ is the CFT Hamiltonian, and
         $C/\delta$ is the Casimir energy. Here we use the same symbol for a defect line and
         for  the corresponding operator. Note that the divergent (or vanishing)  factor $e^{- C/\delta}$ is an overall normalization that
          drops out of  the calculation of   correlation functions.

  \smallskip          
          
      The analysis of \cite{Bachas:2007td} was recently extended 
       to many  free bosons and   fermions   in  reference  \cite{Bachas:2012bj}.  Since the $c=1/2$ CFT is the theory of a free fermion, 
       all we have to do  is  to translate 
     the relevant calculations  of the latter reference to the language of the Ising model. 


 \section{Conformal defects of the Ising model}
 
The critical defect lines of the Ising model can be mapped, using the folding trick,  
to   boundary conditions in the $c=1$ orbifold theory
  \cite{Oshikawa:1996ww,Oshikawa:1996dj}.   The idea is illustrated in figure 2: 
   the $\mathbb{Z}_2$ orbifold  of a free boson on a circle 
    describes the critical line of the Ashkin-Teller model. It reduces to two decoupled Ising models when the radius\footnote{We use the normalization in which the free boson theory is self dual at radius $r=\frac{1}{\sqrt{2}}$.} of the circle is $r=1$  \cite{Ginsparg:1988ui}. 
       Unfolding  converts any boundary condition  of the $r=1$ orbifold to a defect line 
    of the Ising model, and vice versa, whence the equivalence. 
 

\begin{figure}[ht]
\centering
\includegraphics[width=0.85\textwidth]{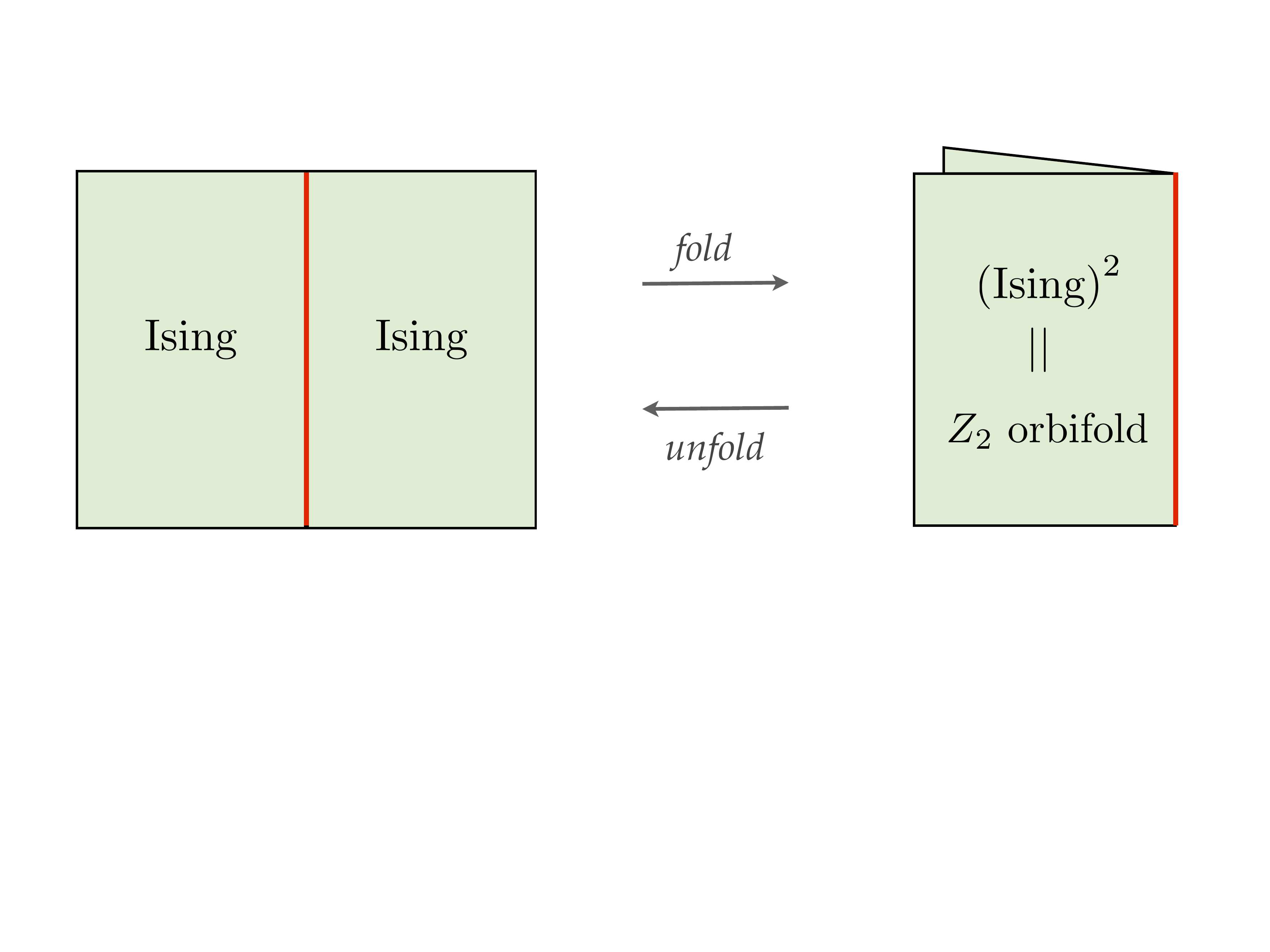} 
\vskip -30mm
\caption{\footnotesize Folding transforms  defect lines (in red) of the critical  Ising model to boundary 
conditions of the $c=1$ $\mathbb{Z}_2$-orbifold theory.  }\label{fig:fold}
\end{figure} 
 \medskip 
 
 As explained in \cite{Oshikawa:1996ww,Oshikawa:1996dj}, see also \cite{Quella2}, 
 the conformal boundary conditions of the orbifold theory  come in two continuous families:   
 \begin{center}
\begin{minipage}{12cm}
\begin{itemize}
\item the Dirichlet condition  $\vert D, \phi_0 \, \rangle\hskip -0.55mm \rangle$ with $\phi_0 \in [ 0, \pi]$, 
 and
\item the Neumann condition $\vert N, \tilde \phi_0 \, \rangle\hskip -0.55mm \rangle$ with $\tilde \phi_0 \in [ 0, \pi/2]$. 
\end{itemize}
\end{minipage}
\end{center}
 In the  language of string theory, $\phi_0$ is the position of a D0-brane on the circle, modulo the $\mathbb{Z}_2$ identification, whereas
 $\tilde\phi_0$ is the Wilson line on a D1-brane, or equivalently
 the position of the dual D0-brane on the dual circle (of radius $\tilde r= 1/2$). 
   Here we have specified the boundary conditions by means of the corresponding boundary states
 $\vert {\cal B}  \, \rangle\hskip -0.55mm \rangle$. 
  Unfolding  converts the  boundary states  of (Ising)$^2$ to  defect operators of the Ising model. 
 \smallskip  
   
  The relation between   $\phi_0$ and the parameter  $b$ of the ``defective"  model  \eqref{Isingc} has been obtained  in
   \cite{Oshikawa:1996ww,Oshikawa:1996dj}  by comparing the CFT spectrum with the exact 
diagonalization  of the transfer matrix 
\cite{Abraham:1989cz,Delfino:1994nr}:\footnote{We have exchanged the role of horizontal and vertical compared to references  \cite{Oshikawa:1996ww,Oshikawa:1996dj}.}
    \be\label{b_phi}
   {\rm tan} (\phi_0 - {\pi\over 4}) = {{\rm sinh} (K_1(1-b)) \over {\rm sinh} (K_1(1+ b))}\ \Longleftrightarrow\
   {\rm cot} \phi_0 =   \,  {{\rm tanh} (b K_1)  \over {\rm tanh} (K_1) }
    \ .   
   \ee
 Note that $\phi_0 = \pi/4$ corresponds to $b=1$, {\it i.e.} to no defect. The corresponding operator is the identity operator. 
 Another special value is $\phi_0 = 3\pi/4$,  which corresponds to $b=-1$. This defect line
  can be removed by flipping the signs of all spins on one side of the defect.
   
\smallskip

Three other special values are $\phi_0 = 0, \pi/2$ and $\pi$, corresponding to $b= \infty, 0$ and $-\infty$ respectively. 
At these special values  the defect line reduces to separate boundary conditions for the two Ising models,  namely\footnote{At
 the two endpoints of the $\phi_0$  interval one actually  finds the sum of two elementary boundary conditions.
 These correspond to the fractional branes sitting at the  fixed points  of the $\mathbb{Z}_2$ orbifold \cite{Pradisi:1988xd,Douglas:1996xg}.}   
\be
(++) \oplus (--)\ , \ \ (ff) \ \  {\rm and} \ \  (+-)\oplus (-+) \ , 
\ee
where $(+), (-), (f) $ denote the three conformal boundary conditions of the Ising model: 
 spin-up, spin-down and free  \cite{Cardy:1989ir}. 

\smallskip
     
 In the infinitely anisotropic limit,  $K_1\to 0$   and ${\rm sinh} (2K_2) \simeq 1/2K_1 \to \infty$,  
 the critical Ising model with a defect line of Dirichlet type can be  described equivalently by the quantum-spin chain  with Hamiltonian
  \cite{Henkel:1988vh}
\be
H_D  =    - \sum_n  \,  h^*    \sigma_n^x  - \sum_{n\not= 0}  \sigma_n^z \sigma_{n+1}^z   -b\,  \sigma_{0}^z \sigma_{1}^z \ , 
\ee
where $h^* = 1$ is the critical value of the transverse magnetic  field.  The defect sits on the link $\langle 01 \rangle$ of the spin chain, and this 
Hamiltonian describes the evolution in the direction parallel (not transverse) to the defect line. The coupling at the defective link is $b = {\rm cot} \phi_0$.  
\smallskip

 In the quantum spin-chain language  one can also describe the
   Neumann family of conformal defects whose Hamiltonian  is  \cite{Oshikawa:1996dj} 
\be
H_N  =   - \sum_n \,  h^*    \sigma_n^x  - \sum_{n\not= 0}  \sigma_n^z \sigma_{n+1}^z   -\tilde b\,  \sigma_{0}^z \sigma_{1}^x \ .  
\ee
Here again  $\tilde b = {\rm cot} \tilde\phi_0$, but one may now   
  restrict   $\tilde b \geq 0$, so that $\tilde \phi_0$ only takes values in the  interval $[0, \pi/2]$. 
 This follows from  the
  automorphism of the Pauli matrices  $(\sigma^x , \sigma^y , \sigma^z) \to (\sigma^x , -\sigma^y , -\sigma^z)$ which flips the sign of
$\tilde b$ while leaving the bulk Hamiltonian unchanged. 

\smallskip
 
 The nature of the Neumann defects is made  transparent by a Kramers-Wannier duality
   of the half-chain  $ n> 0 $. This maps 
  $\sigma_1^x$ to $\mu_1^z$,   
 where $\vec \mu_n $ are the disorder operators, and  the  Neumann defect to  an order-disorder coupling 
 of the two half-chains \cite{Oshikawa:1996dj}.
  When  $\tilde\phi_0 = \pi/4$ we have  $\tilde b=1$, and the Neumann  defect is topological;  it  
  implements the order-disorder duality  in  the $c=1/2$ conformal field theory   \cite{Frohlich:2004ef}.
  At the  endpoints $\tilde\phi_0 = 0, \pi/2$ on the other hand  the  defect reduces to the separate
  boundary conditions
\be
(+f)\oplus (-f) \ \ \ {\rm and} \ \ (f+) \oplus (f-)\ .
\ee

\medskip

Two interesting quantities that characterize all conformal defects are  the ground-state degeneracy  $g$  \cite{Affleck:1991tk}  and
   the reflection coefficient  ${\cal R}$, 
   given by the 2-point function of the energy momentum tensor  \cite{Quella2} 
  \be
 {\cal R} :=  { \langle T_1 \bar T_1 + T_2 \bar T_2 \rangle
 \over
  \langle (T_1 + \bar T_2) (\bar T_1 +  T_2) \rangle }  \ .   
 \ee
Here, $T_1, \bar T_1$ are the components of the energy-momentum tensor  at any point  $z$, while 
 $T_2, \bar T_2$ are evaluated at  the point obtained  by reflection with respect to the defect line.
 For the defects of interest here  one finds:
  \bea
&&\underline{\rm Dirichlet}:  \ \ g=1\ , \ \  {\cal R}  = {\rm cos}^2(2\phi_0) \nonumber  \\
 && \underline{\rm Neumann}:  \ \ g=\sqrt{2}\ , \ \  {\cal R}  = {\rm cos}^2(2\tilde \phi_0) \ . 
 \eea
Note that at $\phi_0  = n\pi/2$,  where the Dirichlet defect  reduces to  totally-reflecting  boundary conditions, the reflection coefficient is ${\cal R}=1$. 
Conversely, at  $\phi_0 = \pi/4$ or $3\pi/4$  the defect is topological and there is no reflection,  ${\cal R}=0$. 
Similar  statements hold for the Neumann defects. 
   
 

  \section{Folding-unfolding  dictionary}
 
   In order to calculate the fusion product defined in \eqref{regularize}  we need to unfold the boundary states 
   of the orbifold theory to defect operators acting on the space of states of the Ising model. 
   The  critical Ising model  is described by a free massless fermion field with components
    \begin{equation}\label{psi}
  (\psi  , \bar \psi ) = \sum_{r} \  ( \psi_r e^{-r(\tau+i \sigma)} \ , \ 
  \bar \psi_r e^{-r(\tau-i\sigma)} )  \ . 
\end{equation} 
Here, $z= \tau+i\sigma$ parametrizes the cylinder $\mathbb{R}\times [0, 2\pi]$, and 
    the Fourier modes satisfy   the canonical anticommutation relations $\{ \psi_r , \psi_s \} =\{ \bar\psi_r , \bar\psi_s \} =  \delta_{r+s, 0}\ .$
    The left  and right components of the energy-momentum tensor are given by
  \be
  T =   - {1\over 2}:\hskip -0.7mm \psi\,  \partial \psi \hskip -0.5mm : \ \ \ {\rm and} \  \ \    \bar T =   - {1\over 2}:\hskip -0.7mm 
  \bar \psi\,  \bar \partial \bar \psi  \hskip -0.5mm : \ , 
  \ee
where $\partial \equiv \partial/\partial z$,  $\bar\partial \equiv \partial/\partial \bar z$, and the double dots stand for  normal ordering. 
\smallskip 

The fermion can be antiperiodic (Neveu-Schwarz)
or periodic (Ramond), and we denote the corresponding ground states by $\vert 0\rangle_{\rm NS}$ and 
$\vert 0  , {\scriptstyle{A}}   \rangle_{\rm R}$, $ {\scriptstyle{A}} = \pm $. The two Ramond ground states
  represent  the Dirac algebra of the zero modes $\psi_0$ and $\bar\psi_0$. 
The Ising CFT  can be obtained from the free fermonic theory by a projection onto  even fermion parity 
 which  acts 
as a chiral projection  on the Ramond ground states. This in particular lifts the ground state degeneracy in the Ramond sector.  
The three primary fields of the Ising model
$\boldsymbol{1}, \boldsymbol{\epsilon}$ and  $\boldsymbol{\sigma}$, 
are mapped by the
 operator-state correspondence    to the states
 $\vert 0\rangle_{\rm NS}$, $\psi_{-1/2}\bar\psi_{-1/2}\vert 0\rangle_{\rm NS}$ and $\vert 0, + \rangle_{\rm R}$, respectively. 
\medskip

Consider now a defect placed on the circle  $\tau = 0$ around the cylinder.  Conformal invariance is tantamount to continuity
of $T-\bar T$. Equivalently, the Fourier modes
     \bea
  \int_0^{2\pi}  {d\sigma \over 2\pi}\,   e^{iN\sigma}\ ( T  - \bar T)  
  \,   =  \,  {1\over 2} \sum_r  ({r} + {N\over 2}) 
   ( :\hskip -0.7mm \psi_{-r} \psi_{N+r} \hskip -0.5mm : + :\hskip -0.7mm \bar\psi_{r} \bar\psi_{-N-r} \hskip -0.5mm :) \ 
    \eea
   on both sides of the defect line have to agree. 
 This  is obviously guaranteed
  by the gluing conditions\,\footnote{The factor of -$i$ ensures that this gluing condition is consistent with the  
   Majorana property $\psi^* = i \bar\psi$ in Euclidean spacetime.}
    \be   \label{glue}
\left(\begin{array}{c} \psi_{-r}  \\
   \,  \\  -  i  \, \bar  \psi_{r}   \end{array}\right) \, {\cal D} \, = \,  {\cal D}\, \Lambda 
   \left(\begin{array}{c} \psi_{-r}  \\
   \,  \\  -  i  \, \bar  \psi_{r}   \end{array}\right) \ , 
\ee
provided  $\Lambda$ is an element of  $O(1,1)$, the group of Lorentz transformations in 1+1 dimensions, {\it i.e.}
$\Lambda^t \eta \Lambda = \eta$ for $\eta = {\rm diag}(1, -1)$.
In the above equation  ${\cal D}$ is the defect operator, and the mode operators acting on the left and right of it come from fields on the left ($\tau<0$) and right ($\tau>0$) of the defect line respectively.
\smallskip

  To relate 
  \eqref{glue} 
  to the boundary states of the previous section  we must fold the half-cylinder  $\tau>0$,  
 so that we now have two fermions at $\tau<0$. Time reflection exchanges  left- and right-movers, 
\beq\label{spell}
\vect{ \psi_r }{  \bar  \psi_{r} }  \to \vect{-i \bar  \psi_{-r} }{ i  \psi_{-r} } \  , 
\eeq
and a little algebra allows us to convert  \eqref{glue}  into a boundary condition  
for the two-fermion theory   \cite{Bachas:2012bj} 
 \beq\label{fermbdgluing2}
\left[\vect{ \psi^1_{r}}{\psi^2_r }  +   i \OOO  \vect{  \bar  \psi^1_{-r} }{  \bar  \psi^2_{-r} }\right] 
\vert {\cal B}  \, \rangle\hskip -0.5mm \rangle\, = \, 0\, ,   
\eeq
where $\OOO$ is the  2$\times$2 rotation  matrix 
 \beq\label{SintermsofO}
\OOO(\Lambda) =\mat{ \Lambda_{12} \Lambda_{22}^{-1}}
{\Lambda_{11}-\Lambda_{12}\Lambda_{22}^{-1}\Lambda_{21}} {\Lambda_{22}^{-1}} {-\Lambda_{22}^{-1} \Lambda_{21}} \, . 
\eeq
Equation \eqref{SintermsofO} maps the Lorentzian group $O(d,d)$ to
 the rotation group $O(2d)$  for any $d$, but we
 will only need it   for $d=1$ here.  

\smallskip 
 
 The group $O(1,1)$ has four  connected components containing 
   the four elements $\Lambda = {\rm diag}(\pm 1, \pm 1)$ respectively. 
   Due to the 
    projection onto even fermion parity,   $\Lambda$ and $-\Lambda$ describe  equivalent gluings
  so that there are only two continuous families of gluing conditions. 
    The ones with  ${\rm det} \Lambda= +1$ correspond to the Dirichlet boundary conditions in the orbifold theory,
    {\it i.e.} to the (anti-)ferromagnetic defect lines,
   whereas  the ones with ${\rm det} \Lambda= -1$ correspond to the Neumann  boundary conditions, {\it i.e.} to the order-disorder defect lines.

\smallskip

 To establish the exact dictionary, we first use  \eqref{SintermsofO} to relate the gluing matrix $\Lambda$
  (for ${\rm det}\Lambda = +1$)   to the following rotation matrix: 
\bea 
\Lambda  =   \mat{{\rm cosh}\gamma}{{\rm sinh}\gamma}{{\rm sinh}\gamma}{{\rm cosh}\gamma}\  \leftrightarrow \ 
\OOO  = \mat{{\rm cos}(2 \phi_0)} {{\rm sin}(2 \phi_0)} {{\rm sin}(2 \phi_0)} {-{\rm cos}(2 \phi_0)}
 \ , 
\eea
where 
\beq\label{219}
 {\rm cos}(2 \phi_0) =  {\rm tanh}\gamma \  \ \Longleftrightarrow\ \ 
  e^{ \gamma} = {\rm cot}\phi_0  \ .  
\eeq
The  bosonization formulae $ \psi^1+ i\psi^2 = {\rm exp} (2\int\partial\phi)$ and
$ \bar\psi^2+ i\bar\psi^1 = {\rm exp} (2\int\bar\partial\phi)$, 
   and the boundary condition \eqref{fermbdgluing2} allow us to identify the angle $\phi_0$
with the D0-brane position on the orbifold space. As $\gamma$ ranges from $\infty$ to $-\infty$, $\phi_0$  takes  values in $[0, \pi/2]$. 
However, gluing in the Ramond sector involves the spinor representation  $S(\OOO)$ of the orthogonal group $O(2)$. This effectively doubles the range of $\phi_0$, in agreement with the discussion of section 3: the defects with $\phi_0\in [0,\pi/2]$ correspond to defects with $a=\boldsymbol{1}$, whereas the ones with
$\phi_0\in [\pi/2,\pi]$ correspond to defects with $a=\boldsymbol{\epsilon}$.

\smallskip

Combining  equations  \eqref{219} and  \eqref{b_phi} yields relation \eqref{relnt} between the Ising model parameter $b$ and the hyperbolic angle $\gamma$ quoted in the introduction.

\smallskip
 The gluing conditions \eqref{glue} with ${\rm det}\Lambda = -1$ fold to boundary gluings  
 \eqref{fermbdgluing2},  with the following O(2) matrix:  
\bea\label{torth}
\Lambda  =
\mat{{\rm cosh}\tilde\gamma}{-{\rm sinh}\tilde\gamma}{{\rm sinh}\tilde\gamma}{-{\rm cosh}\tilde\gamma} \ 
 \  \leftrightarrow \ 
  \OOO  = \mat{{\rm cos}(2\tilde  \phi_0)} {{\rm sin}(2\tilde  \phi_0)} {-{\rm sin}(2\tilde \phi_0)} {{\rm cos}(2\tilde \phi_0)}\ , 
 \eea
where $\tilde\gamma$ is related to  $\tilde\phi_0$  as in \eqref{219}.  Since transformations with ${\rm det}\Lambda = -1$
  flip the chirality of  $O(1,1)$ spinors, such defect operators
 cannot act consistently in the Ramond sector \cite{Bachas:2012bj}. As a result,  
  one may restrict 
 $\tilde\phi_0\in [0, \pi/2]$.
\medskip

 The boundary states obeying conditions \eqref{fermbdgluing2} were constructed explicitly and unfolded into defect operators  
 in    reference  \cite{Bachas:2012bj}. In a somewhat  elliptical notation they read: 
 \bea
{\cal D}^{\pm}  \   = \  {\cal T}( \prod_{r>0} e^{-i  \sum_{j,k}  \OOO_{jk}
 \psi^j_{-r}\bar\psi^k_{-r}}  )   {1\over 2}   \Bigl[   \mathbb{I}_0^{\rm NS} \,  \pm  
   \,  {1\over \sqrt{{\rm cosh}\gamma} } \,   \mathbb{I}_0^{\rm R} \, S(\Lambda)    \Bigr]   
  \ + \ (\Lambda\mapsto -\Lambda)\   \nonumber
\eea
\vskip -3mm
\bea
{\rm and} \ \ \  \ \ 
\tilde {\cal D}  = \,   \  {\cal T}( \prod_{r>0} e^{-i  \sum_{j,k}   \OOO_{jk}
 \psi^j_{-r}\bar\psi^k_{-r}}  )   {1\over\sqrt{2}} \,  \mathbb{I}_0^{\rm NS}  \ + \ (\Lambda\mapsto -\Lambda)\ , \nonumber
\eea
where
\be 
 \mathbb{I}_0^{\rm NS} = \vert 0  \rangle_{\rm NS} \ _{\rm NS}\langle 0 \vert \qquad  {\rm and} 
\qquad  \mathbb{I}_0^{\rm R}  = \sum_{\scriptstyle{A}}\vert 0, {\scriptstyle{A}}     \rangle_{\rm R}  \ _{\rm R}\langle 0, {\scriptstyle{A}}   \vert 
\ee
 are the identity operators in the ground-state sectors. ${\mathcal D}^\pm$ are the defect operators for ${\rm det}\Lambda = +1$
 and $\tilde{\mathcal D}$ the ones for ${\rm det}\Lambda = -1$.
   Furthermore  $\OOO$ is  the orthogonal matrix  given in
  \eqref{SintermsofO}, and the
  oscillator frequencies $r$ run over the positive integers or half-integers in the
  periodic, respectively antiperiodic  sectors. Finally ${\cal T}$ is the time-reversal operation \eqref{spell}
  which acts only on the  $j=2$ fermions, {\it i.e.} on the copy of the Ising CFT that is being unfolded. 
  
  The meaning of the above formulae is as follows:
  expand the exponentials, apply the operation ${\cal T}$, and act by the fermion modes with index $j=1$ on the left and those with index $j=2$ on the right 
  of  the ground-state isomorphisms $ \mathbb{I}_0^{\rm NS}$ or $ \mathbb{I}_0^{\rm R}$. 
\smallskip

 In the notation of the  introduction  we have the following correspondence between defect lines and operators: 
 \bea
  (\boldsymbol{1}, \Lambda) \mapsto {\cal D}^+(\Lambda)\ , \;
  (\boldsymbol{\epsilon}, \Lambda) \mapsto {\cal D}^-(\Lambda)\qquad
 & {\rm for}& \qquad {\rm det} \Lambda=1\nonumber\\
{\rm and}\qquad   (\boldsymbol{\sigma}, \Lambda) \mapsto \tilde {\cal D}(\Lambda)\qquad&{\rm for}&\qquad {\rm det}\Lambda=-1\ .   
 \eea
 The translation in the language of the Ising model was given in table 1.
  
   The order-disorder defect $\tilde{\cal D}$ has no Ramond  component.  Since the spin operator is 
   in the Ramond sector, we conclude  that there is 
   no  correlation between spin operators on either side
   of such  defect  lines.
  
  
  \section{Computing the fusion}
  
 Having constructed the defect operators, we can now compute the fusion of defects as defined in \eqref{regularize}. This was done  (for any  number of fermion fields)  in \cite{Bachas:2012bj}. 
  We will recall the main steps
 of this  calculation here.
 \smallskip
 
  Note first  that all  defect operators are (sums of)   products of the  form 
    \be 
{\cal D}=    \,  {\cal D}_0  \,  \prod_{r>0} {\cal D}_r   \ , 
\ee 
 where ${\cal D}_r $ only involves the fermion modes  $\psi_{\pm r}^j $ and $\bar\psi_{\pm r}^j $, while
   ${\cal D}_0$ gives the action of the defect operator  on the ground states.
    The operators ${\cal D}_r$ for different $r$  commute, so their order is irrelevant.  Hence, in evaluating 
    the product in \eqref{regularize}, 
    we may consider each term   ${\cal D}_r e^{-\delta {\cal H}} {\cal D}^\prime_r$ separately. 
  \smallskip 
    
    We  use the
    label $j=1,2,3$ to denote the fermion field in the region on the left of both defects, in the region between the two defects, and
    finally the region on the right (see figure 2).   Thus, the  operator 
    ${\cal D}$ involves the fermions $j=1,2$ and ${\cal D}^\prime$ the fermions $j=2,3$.  Now the idea is to anticommute the
    common fermions,  $j=2$,  so as to bring all positive-frequency (annihilation) operators to the right  of
    all negative-frequency (creation) operators.  The result can then be easily evaluated, since
    it is  sandwitched between ground states of theory 2. One ends up with an expression  that only involves the fermions $j=1,3$,
    which are spectators in this rearrangement. 
\smallskip

  To perform this calculation we  use  the following identities:  
   \be\label{iddf1}
 e^{  \chi \psi_{r}}  \, f(\psi_{-r}) \, =   \, f(\psi_{-r}+ \chi)\,  e^{\chi \psi_{r}}  \ , 
 \ee 
valid for any function $f$ and any operator $\chi$  that  anticommutes  with the $\psi_{\pm r}$,   
and  
\be
\langle  0 \vert  \  e^{ u\,    \psi_{r}   \bar \psi_r } \,  e^{u^\prime \psi_{-r}   \bar  \psi_{-r} }  
 =    (1- u\, u^\prime  )\, \langle  0 \vert   \, {\rm exp}\left( {u\over 1-u\, u^\prime} \,  \psi_{r}   \bar \psi_{r}\right) \ ,  
\ee
where $u, u^\prime$ are c-numbers. Consider two defects with gluing matrices $\Lambda$ and $\Lambda^\prime$  
and corresponding orthogonal matrices $\OOO$ and $\OOO^\prime$. 
Using the above  identities leads  after some tedious  algebra to  \cite{Bachas:2012bj}
\be
{\cal D}_r e^{-\delta {\cal H}} {\cal D}^\prime_r \ =\  (1- e^{-2\delta r} \OOO_{11}^\prime \OOO_{22} )\, 
 {\cal T}( e^{-i  \sum_{j,k}   \OOO_{jk}^{\prime\prime}(e^{-2\delta r}) 
 \psi^j_{-r}\bar\psi^k_{-r}} ) \ , 
\ee
where  
the   indices $j,k$ take the values $1$ and $3$ [the fermions $\psi^2$ and $\bar\psi^2$
have been  integrated out]. Moreover, the matrix $\OOO^{\prime\prime}(x)$ is given by
 \be\label{compmat}
{ \scriptsize
  \OOO^{\prime\prime}(x)=\mat{\OOO_{11}+x^2\OOO_{12}(1-x^2\OOO^\prime_{11}\OOO_{22})^{-1}\OOO^{\prime}_{11}\OOO_{21}}
  {x\OOO_{12}(1-x^2\OOO^{\prime}_{11}\OOO_{22})^{-1}\OOO^{\prime}_{12}}
 {x\OOO^{\prime}_{21}(1-x^2\OOO_{22}\OOO^{\prime}_{11})^{-1}\OOO_{21}}
 {\OOO^{\prime}_{22}+x^2\OOO^{\prime}_{21}(1-x^2\OOO_{22}\OOO^{\prime}_{11})^{-1}\OOO_{22}\OOO^{\prime}_{12}} \, .
 }
 \ee
 In the limit $\delta\to 0$, $\OOO^{\prime\prime}(e^{-2\delta r})$ converges to the orthogonal matrix 
 corresponding to the product $\Lambda\Lambda^\prime$ of gluing matrices. However the infinite product of numerical factors  $\prod_r(1- e^{-2\delta r} \OOO_{11}^\prime \OOO_{22} )$  does not converge nicely in the limit. 
%
%
 Its behavior can be computed with the help of  the following Euler-Maclaurin expansions \cite{Bachas:2012bj}:
\begin{eqnarray}
\prod_{r\in \mathbb{N}+1/2}  (1- e^{-2\delta r}\, \OOO_{11}^\prime\OOO_{22}) &\simeq&  
   e^{C/\delta}  \, (1 + o(\delta))  \qquad {\rm and} \nonumber\\
 \prod_{r\in \mathbb{N}}  (1- e^{-2\delta r}\, \OOO_{11}^\prime\OOO_{22}
) &\simeq&  
  (1-\OOO_{11}^\prime\OOO_{22})^{-1/2} \,  e^{C/\delta} \, (1 + o(\delta))  \ ,  \nonumber \\
  {\rm with} &&C = \int_0^\infty  dx  \, {\rm log} (1 -e^{-2x}\, \OOO_{11}^\prime\OOO_{22}  ) \ . 
\end{eqnarray}
In the antiperiodic (Neveu-Schwarz) sector, the exponential singularity is exactly removed by the counterterm in the definition 
\eqref{regularize} of fusion, whereas in the periodic (Ramond) sector there is a left-over factor 
 \be
 (1-\OOO_{11}^\prime\OOO_{22})^{-1/2}= (1+ {\rm tanh}\gamma \, {\rm tanh}\gamma^\prime)^{-1/2}  =  \left( { {\rm cosh}\gamma \, {\rm cosh}\gamma^\prime \over
 {\rm cosh}(\gamma +  \gamma^\prime) }  \right)^{1/2}\ .  
 \ee
 This factor is essential for the fusion to produce a properly normalized defect operator in the Ramond sector. 
 Here we assumed ${\rm det} \Lambda
 = {\rm det} \Lambda^\prime = +1$, which is sufficient, because 
  only the Dirichlet defects have a non-trivial component in the Ramond sector.
  \smallskip

The rest of the calculation is straightforward  and leads  to the  following fusion of defects:
\be
{\cal D}^+ (\Lambda) \star {\cal D}^\pm(\Lambda^\prime)  = {\cal D}^\pm(\Lambda\Lambda^\prime)\ ,
 \qquad {\cal D}^- (\Lambda) \star {\cal D}^\pm(\Lambda^\prime) = {\cal D}^\mp(\Lambda\Lambda^\prime)\ , \nonumber
\ee
\be
{\cal D}^\pm(\Lambda)  \star \tilde {\cal D}(\Lambda^\prime)  =  \tilde {\cal D}(\Lambda\Lambda^\prime) \ ,  \qquad
 \tilde {\cal D}(\Lambda) \star \tilde {\cal D}(\Lambda^\prime)  = {\cal D}^+(\Lambda\Lambda^\prime) + {\cal D}^-(\Lambda\Lambda^\prime)\ . 
\ee
 Note that the composition of the fermion-gluing conditions  \eqref{glue} is classical.  In the quantum theory this  is superposed
 with the Verlinde  algebra of the Ising model, as mentioned in the introduction. 
\smallskip    
 
  The above defects exhaust the universality classes of Ising defects with finite $g$-factor.  The  $c=1$ circle CFT   has extra conformal boundary states
  at rational multiples of the (self-dual) radius  of the circle theory, {\it i.e.} at  $r= p/(q\sqrt{2})$ \cite{GR}. 
  At a special point in their moduli space these states reduce to a superposition of $q$  equally spaced Dirichlet branes 
   $\vert D, \phi_0 \, \rangle\hskip -0.55mm \rangle$. The 
    radius $r=1$ that interests us here is however irrational. If consistent    boundary states  still exist \cite{Janik},
    they should correspond to smeared-out limits of infinitely many Dirichlet branes, and hence have a divergent $g$ factor. 
    We did  not consider such  boundary conditions here. 

 The stability of the defect lines considered in this paper has been analyzed  in reference \cite{Oshikawa:1996dj}. The Neumann defects preserve the global
  $\mathbb{Z}_2 \times \mathbb{Z}_2$ symmetry under reversal of the spins on either side of the defect line, while the more stable
  Dirichlet defects only preserve the diagonal $\mathbb{Z}_2$. Perturbations that break the symmetry completely drive the system to
  the totally-reflecting Dirichlet conditions at $\phi_0 = 0, \pi$. Similar considerations should apply to
  the stability of the fusion product.

\vskip 1cm
\noindent {\bf Acknowledgements}
\vskip 2mm
We thank   Denis Bernard  for a conversation,  and the referee of   \cite{Bachas:2012bj} for encouraging us to 
translate the results of this reference  in the language of the  Ising model. We also acknowledge useful discussions with  the participants of
the Hamburg Workshop on ``Field Theories with Defects". 
 


\begin{thebibliography}{99}

\bibitem{Onsager:1943jn}
  L.~Onsager,
  ``Crystal statistics. 1. A Two-dimensional model with an order disorder transition,''
  Phys.\ Rev.\  {\bf 65} (1944) 117.

\bibitem{Binder}
 K. Binder, in {\it Critical behavior at surfaces, Phase transitions and critical 
phenomena} vol. 8, edited by C. Domb and J. Lebowitz (Academic Press, 
London, 1983). 


\bibitem{Henkel:1988vh}
  M.~Henkel, A.~Patkos and M.~Schlottmann,
  ``The Ising Quantum Chain With Defects. 1. The Exact Solution,''
  Nucl.\ Phys.\ B {\bf 314} (1989) 609.

\bibitem{Abraham:1989cz}
  D.~B.~Abraham, L.~F.~Ko and N.~M.~Svrakic,
  ``Transfer Matrix Spectrum For The Finite Width Ising Model With Adjustable Boundary Conditions: Exact Solution,''
   J.\ Stat.\ Phys. {\bf 56} (1989)  563.


\bibitem{Oshikawa:1996ww}
  M.~Oshikawa and I.~Affleck,
  ``Defect lines in the Ising model and boundary states on orbifolds,''
  Phys.\ Rev.\ Lett.\  {\bf 77} (1996) 2604
  [hep-th/9606177].

\bibitem{Oshikawa:1996dj}
  M.~Oshikawa and I.~Affleck,
  ``Boundary conformal field theory approach to the critical two-dimensional Ising model with a defect line,''
  Nucl.\ Phys.\ B {\bf 495} (1997) 533
  [cond-mat/9612187].

\bibitem{Bachas:2007td}
  C.~Bachas and I.~Brunner,
  ``Fusion of conformal interfaces,''
  JHEP {\bf 0802} (2008) 085
  [arXiv:0712.0076 [hep-th]].

\bibitem{Verlinde:1988sn} 
  E.~P.~Verlinde,
  ``Fusion Rules and Modular Transformations in 2D Conformal Field Theory,''
  Nucl.\ Phys.\ B {\bf 300}, 360 (1988).

\bibitem{Petkova:2000ip}
  V.~B.~Petkova and J.~B.~Zuber,
  ``Generalized twisted partition functions,''
  Phys.\ Lett.\ B {\bf 504} (2001) 157
  [hep-th/0011021].

\bibitem{Bachas:2012bj}
  C.~Bachas, I.~Brunner and D.~Roggenkamp,
  ``A worldsheet extension of O(d,d:Z),''
  JHEP {\bf 1210} (2012) 039
  [arXiv:1205.4647 [hep-th]].

\bibitem{Callan:1987px}
  C.~G.~Callan, Jr., C.~Lovelace, C.~R.~Nappi and S.~A.~Yost,
  ``Adding Holes and Crosscaps to the Superstring,''
  Nucl.\ Phys.\ B {\bf 293} (1987) 83.



\bibitem{Bachas:2001vj}
  C.~Bachas, J.~de Boer, R.~Dijkgraaf and H.~Ooguri,
  ``Permeable conformal walls and holography,''
  JHEP {\bf 0206} (2002) 027
  [hep-th/0111210].
 
\bibitem{Mikhailov:2007eg}
  A.~Mikhailov and S.~Schafer-Nameki,
  ``Algebra of transfer-matrices and Yang-Baxter equations on the string worldsheet in AdS(5) x S(5),''
  Nucl.\ Phys.\ B {\bf 802} (2008) 1
  [arXiv:0712.4278 [hep-th]].
 
\bibitem{Benichou:2010ts}
  R.~Benichou,
  ``Fusion of line operators in conformal sigma-models on supergroups, and the Hirota equation,''
  JHEP {\bf 1101} (2011) 066
  [arXiv:1011.3158 [hep-th]].

\bibitem{Ginsparg:1988ui}
  P.~H.~Ginsparg,
  ``Applied Conformal Field Theory,''
  hep-th/9108028.
 
 
\bibitem{Quella2}
  T.~Quella, I.~Runkel and G.~M.~T.~Watts,
  ``Reflection and Transmission for Conformal Defects,''
  JHEP {\bf 0704} (2007) 095
  [arXiv:hep-th/0611296].

\bibitem{Delfino:1994nr}
  G.~Delfino, G.~Mussardo and P.~Simonetti,
  ``Scattering theory and correlation functions in statistical models with a line of defect,''
  Nucl.\ Phys.\ B {\bf 432} (1994) 518
  [hep-th/9409076].



\bibitem{Cardy:1989ir}
  J.~L.~Cardy,
  ``Boundary Conditions, Fusion Rules and the Verlinde Formula,''
  Nucl.\ Phys.\ B {\bf 324} (1989) 581.
  
\bibitem{Pradisi:1988xd}
  G.~Pradisi and A.~Sagnotti,
  ``Open String Orbifolds,''
  Phys.\ Lett.\ B {\bf 216} (1989) 59.

\bibitem{Douglas:1996xg}
  M.~R.~Douglas,
  ``Enhanced gauge symmetry in M(atrix) theory,''
  JHEP {\bf 9707} (1997) 004
  [hep-th/9612126].




\bibitem{Frohlich:2004ef}
  J.~Fr\"ohlich, J.~Fuchs, I.~Runkel and C.~Schweigert,
  ``Kramers-Wannier duality from conformal defects,''
  Phys.\ Rev.\ Lett.\  {\bf 93} (2004) 070601
  [cond-mat/0404051].
  
  

\bibitem{Affleck:1991tk}
  I.~Affleck and A.~W.~W.~Ludwig,
  ``Universal noninteger 'ground state degeneracy' in critical quantum systems,''
  Phys.\ Rev.\ Lett.\  {\bf 67} (1991) 161.

 \bibitem{GR}
  M.~R.~Gaberdiel and A.~Recknagel,
  ``Conformal boundary states for free bosons and fermions,''
  JHEP {\bf 0111} (2001) 016
  [hep-th/0108238].

\bibitem{Janik}
  R.~A.~Janik,
  ``Exceptional boundary states at c=1,''
  Nucl.\ Phys.\ B {\bf 618} (2001) 675
  [hep-th/0109021].


  
  
  
\end{thebibliography}
 \end{document}